\documentclass[12pt,british,english,american]{article}
\usepackage[T1]{fontenc}
\usepackage[latin1]{inputenc}
\usepackage{a4wide}
\usepackage{graphicx}
\usepackage{setspace}
\makeatletter

\providecommand{\tabularnewline}{\\}

\usepackage{babel}
\makeatother
\begin{document}

\title{The measurements of 2200 ETL9351 type photomultipliers for the Borexino
experiment with the photomultiplier testing facility at LNGS.}

\maketitle
\begin{center}{\large A.Ianni}%
\footnote{I.N.F.N. Laboratorio Nazionale del Gran Sasso, SS 17 bis Km 18+910,
I-67010 Assergi(AQ), Italy%
}{\large , P.Lombardi}%
\footnote{Dipartimento di Fisica Universitá and I.N.F.N. sez. di Milano, Via
Celoria, 16 I-20133 Milano, Italy%
}{\large , G.Ranucci$^{2}$, and O.Smirnov}%
\footnote{Corresponding author: Joint Institute for Nuclear Research, 141980
Dubna, Russia. E-mail: osmirnov@jinr.ru;smirnov@lngs.infn.it%
}\end{center}{\large \par}

\begin{abstract}
The results of tests of more than 2200 ETL9351 type PMTs for the Borexino
detector with the PMT test facility are presented. The PMTs characteristics
relevant for the proper detector operation and modeling are discussed
in detail. 
\end{abstract}

\newpage
\section{Introduction}

The Borexino detector is a large volume liquid scintillator detector
designed especially for detection of Solar $^{7}Be$ neutrino \cite{Borex}.
The scintillation from the recoil electrons is registered by 2200
8'' PMTs surrounding the detector's active volume. All PMTs before
installation in the detector were carefully tested in the specially
designed test facility. The technical description of the PMT test
facility at Gran Sasso for the Borexino experiment is presented in
another paper \cite{TestFacility}. Here we discuss in detail the
most important PMT characteristics measured during the tests. The
measured characteristics split naturally into 4 classes:

\begin{enumerate}
\item Dark noise measurements (using scalers);
\item Charge spectrum measurements (using charge amplitude-to-digital converters,
ADC);
\item Transit time timing measurements on the short time scale of 100 ns
(using time-to-digital converters (TDC) with 250 ps resolution);
\item Afterpulses timing measurements on the long time scale of 30 $\mu$s
(using Multihit TDCs with 1 ns resolution).
\end{enumerate}
In this paper the detailed description of the results of measurements
for each of the 4 classes is given. The set of characteristics for
the PMT, which passed the acceptance test was put into the database
prepared for the experiment, but because of the huge amount of PMTs
used in the detector, the straightforward use of the database for
the detector's modeling would slow down the calculations. Therefore,
for the purpose of the detector modeling, the average characteristics
of the PMTs for each class of measurement have been determined.

\section{Dark noise }

\begin{figure*}
\begin{center}\includegraphics[%
  width=1.0\textwidth,
  height=0.60\textwidth]{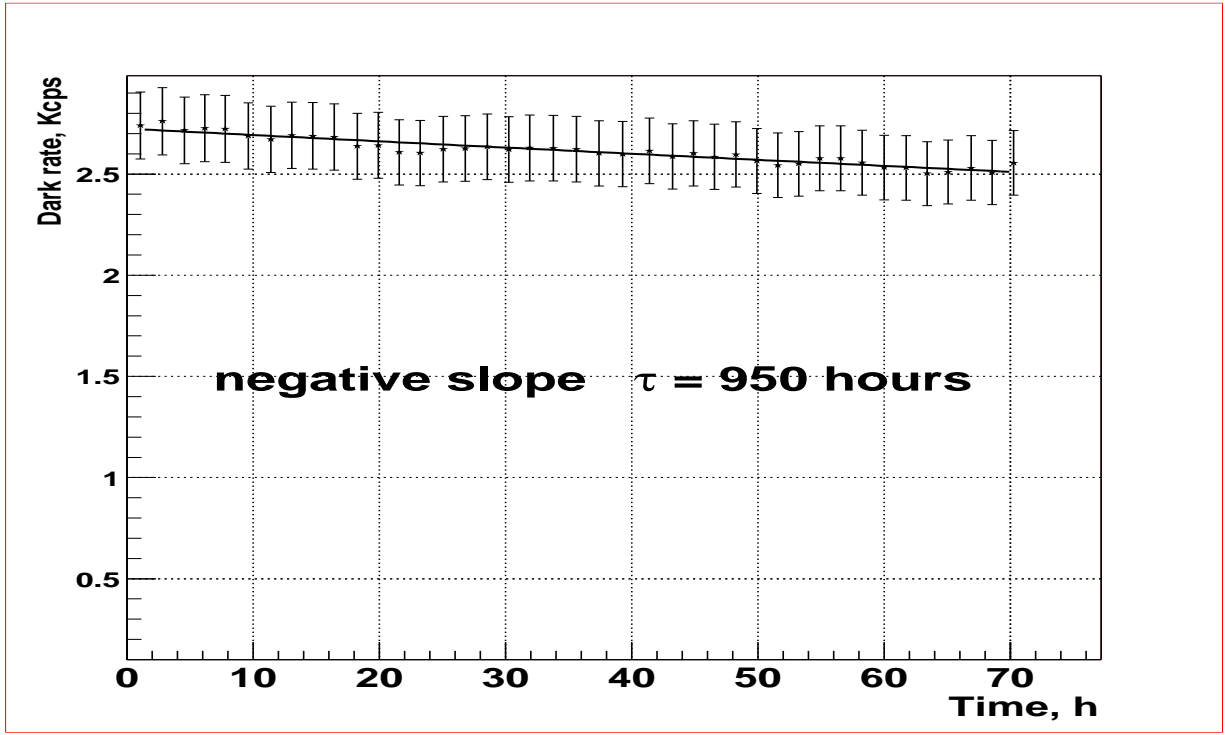}\end{center}

\caption{\label{Figure:GoodPMT1}Dark rate of the ETL ser.No 5797. }
\end{figure*}

In order to minimize the probability of the random trigger in the
detector, the PMTs with a high dark noise rate were rejected. Another
drawback of the high dark noise rate is the shift of the energy scale
of the detector by the amount of the random charge collected during
the event window. The technical specification for the factory was
set at the maximum level of 20000 counts per second (cps). Tests showed
that the initial dark rate is decreasing exponentially during 3-4
hours, then it is still decreasing but much slower, with a characteristic
time of up to 50 days. The typical example of the dark rate behaviour
during a 3 days run is shown in Fig.\ref{Figure:GoodPMT1}. It should
be noted that this is mainly thermoelectron noise and correlated with
it ionic/dynodes afterpulsing (at the level of $\simeq5\%$). 

The other components of the dark rate, namely signals induced by high
energy cosmic rays and by natural radioactivity, are negligible. Furthermore,
at the underground site of the Gran Sasso laboratory (built at a depth
of 3500 meters of water equivalent) the cosmic muon flux is reduced
by a factor $10^{6}$. As concerning the natural radioactivity of
the PMT components, very strong limits on the content of the radioactive
elements from the $U-Th$ chains, and potentially very dangerous $^{40}K$
in the glass of the PMT bulb, were set for the manufacturer. 

\begin{figure*}
\begin{center}\includegraphics[%
  width=1.0\textwidth,
  height=0.60\textwidth]{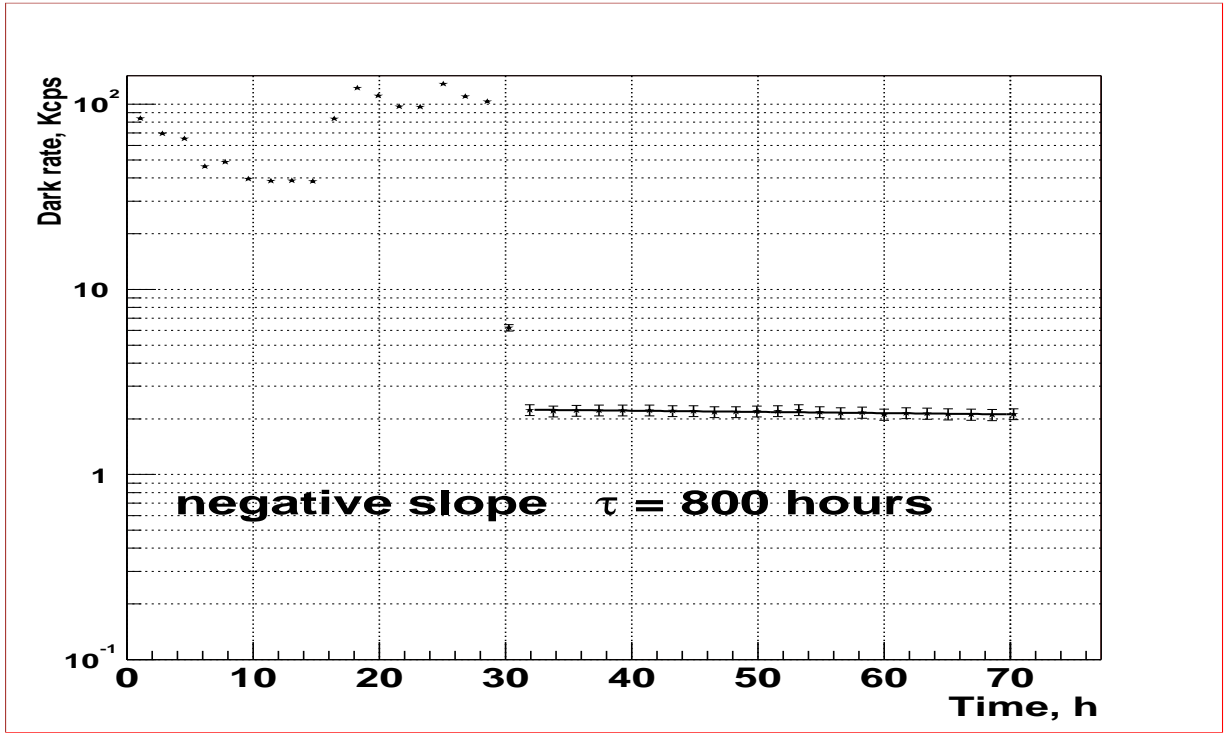}\end{center}

\caption{\label{Figure:BigDarkRateVariations}Dark rate of the ETL ser.No
5832. }
\end{figure*}

\begin{figure*}
\begin{center}\includegraphics[%
  width=1.0\textwidth,
  height=0.50\textwidth]{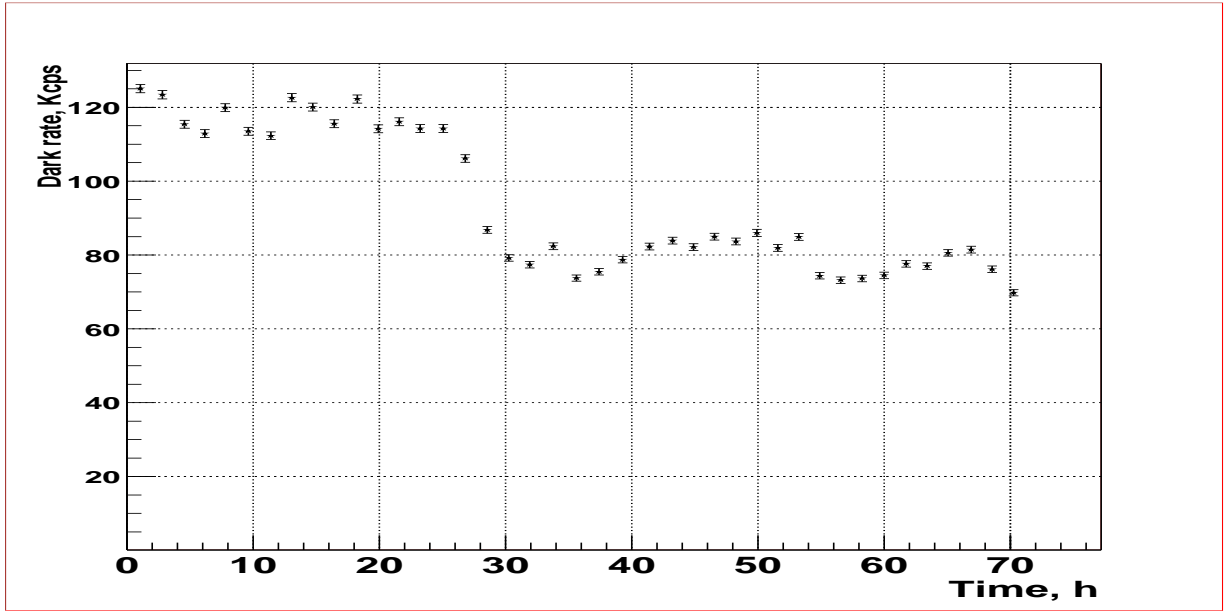}\end{center}

\caption{\label{Figure:DarkRateRejected}Dark rate of the ETL ser.No 5812. }
\end{figure*}

Some of the PMTs demonstrate a very rapid decrease of dark rate, an
example of such behaviour is demonstrated in Fig.\ref{Figure:BigDarkRateVariations}.
The dark rate of the PMT has been changed by one order of magnitude
after 1 day under dark conditions. Very probably this behaviour reflects
some constructive peculiarities of the PMT. Usually, the dark rate
after 3-4 hours was small enough to allow the starting of measurements.
PMTs with a high dark rate noise were kept in the dark for much longer
periods in order to exclude the possibility of rejection of devices
with acceptable dark rate in stable regime. It should be noted that
the dark rate measured at the test facility is usually a factor 2-3
higher than the dark rate at operating conditions in the detector,
due to the much higher time of deexcitation and lower temperature
at the underground site. An example of the dark rate behaviour in
time for a PMT that does not demonstrate dark rate stabilization is
presented in Fig.\ref{Figure:DarkRateRejected}.

\begin{figure*}
\begin{center}\includegraphics[%
  width=1.0\textwidth,
  height=0.50\textwidth]{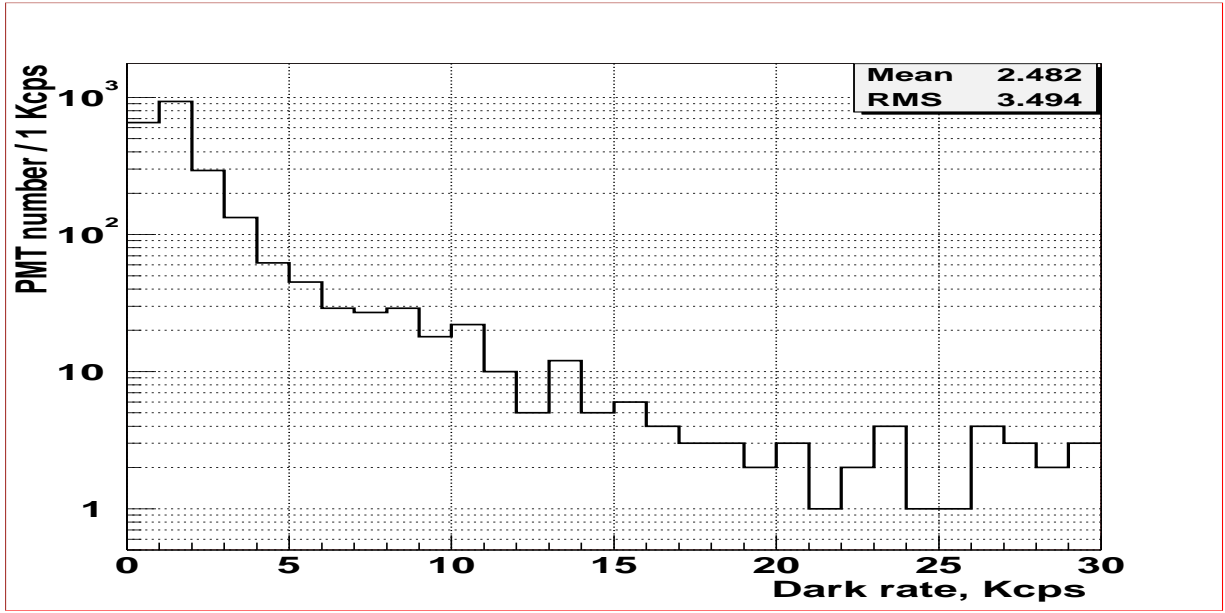}\end{center}

\caption{\label{Figure:fdarkStat}Statistics of the dark rate}
\end{figure*}

The dark rate of each PMT has been measured every 10 minutes during
the test. The results for 2000 PMTs are presented in Fig.\ref{Figure:fdarkStat}. 

It is a well-known fact that the dark rate of the PMTs shows significant
deviations from Poisson statistics. In order to estimate the stability
of the dark rate count we used the excess variation over the Poisson
statistics:

\[
S=\frac{\sqrt{<f_{dark}^{2}>-<f_{dark}>^{2}}}{<f_{dark}>}.\]

In the case of normal (or Poissonian) statistics of the dark rate
measurements the variable $S$ should be close to unity. For the PMTs
with a low dark rate (about 1 Kcps) left for a long period (at least
1 week) under dark conditions the stability $S$ was in a range of
1.2-1.5. The effect of the very slow decrease of the dark rate can
be due to the slow variation of the average ambient temperature (for
the same reason the $S$ factor is greater than 1 even for the very
stable PMTs). 

\begin{figure*}
\begin{center}\includegraphics[%
  width=1.0\textwidth,
  height=0.33\textwidth]{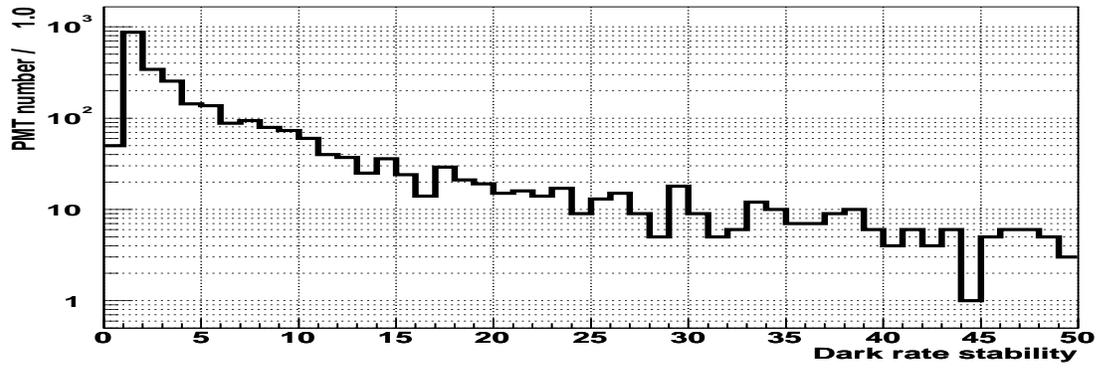}\end{center}

\begin{center}\includegraphics[%
  width=1.0\textwidth,
  height=0.33\textwidth]{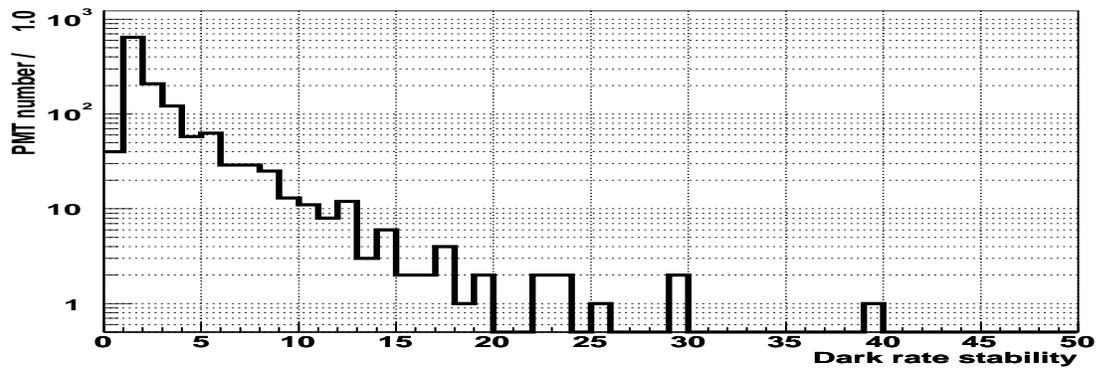}\end{center}

\begin{center}\includegraphics[%
  width=1.0\textwidth,
  height=0.33\textwidth]{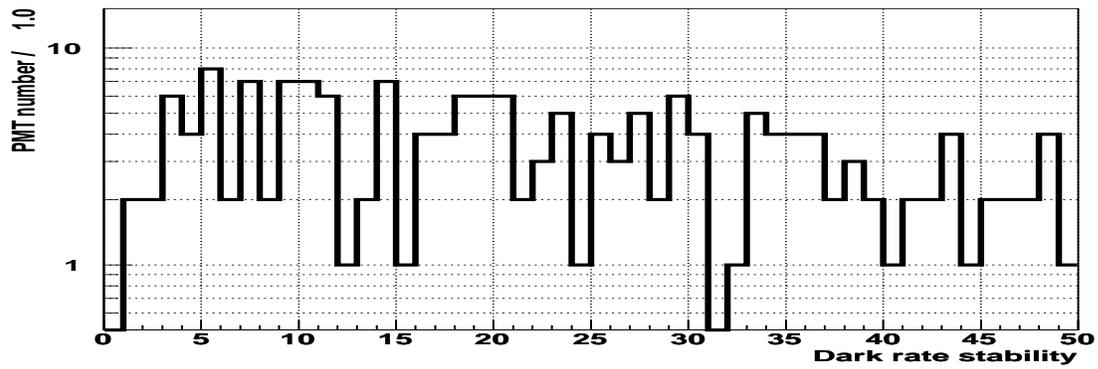}\end{center}

\caption{\label{Figure:FDarkStability}From up to down: stability, S, of the
dark rate for all PMTs, S for the PMTs with low dark rate f<1 Kcps,
and for the PMTs with high dark rate f>10 Kcps. }
\end{figure*}

In the bulk measurements the PMTs were kept in the dark only for the
period necessary for the decay of the fast component of the dark rate
(3-4 hours). The decrease of the dark rate during the measurements
was the main source of the bigger values of the parameter $S$. The
statistics of the parameter $S$ over the measured sample of 2300
PMTs is shown in Fig.\ref{Figure:FDarkStability} (upper plot). One
can see that high values of $S$ , up to 50 and even more, were observed.
A strong correlation between the PMT dark rate and the parameter $S$
exists. The lower plots in Fig.\ref{Figure:FDarkStability} represent
parameter $S$ for the PMTs with low dark rate $f<1$ Kcps, and for
the PMTs with high dark rate $f>10$ Kcps. It is clearly seen the
increase of the relative variance of the dark rate for the PMTs with
higher dark rate. 

\section{Charge spectrum}

\subsection{Charge spectrum structure}

\begin{figure*}
\begin{center}\includegraphics[%
  width=0.80\textwidth,
  height=0.40\textwidth]{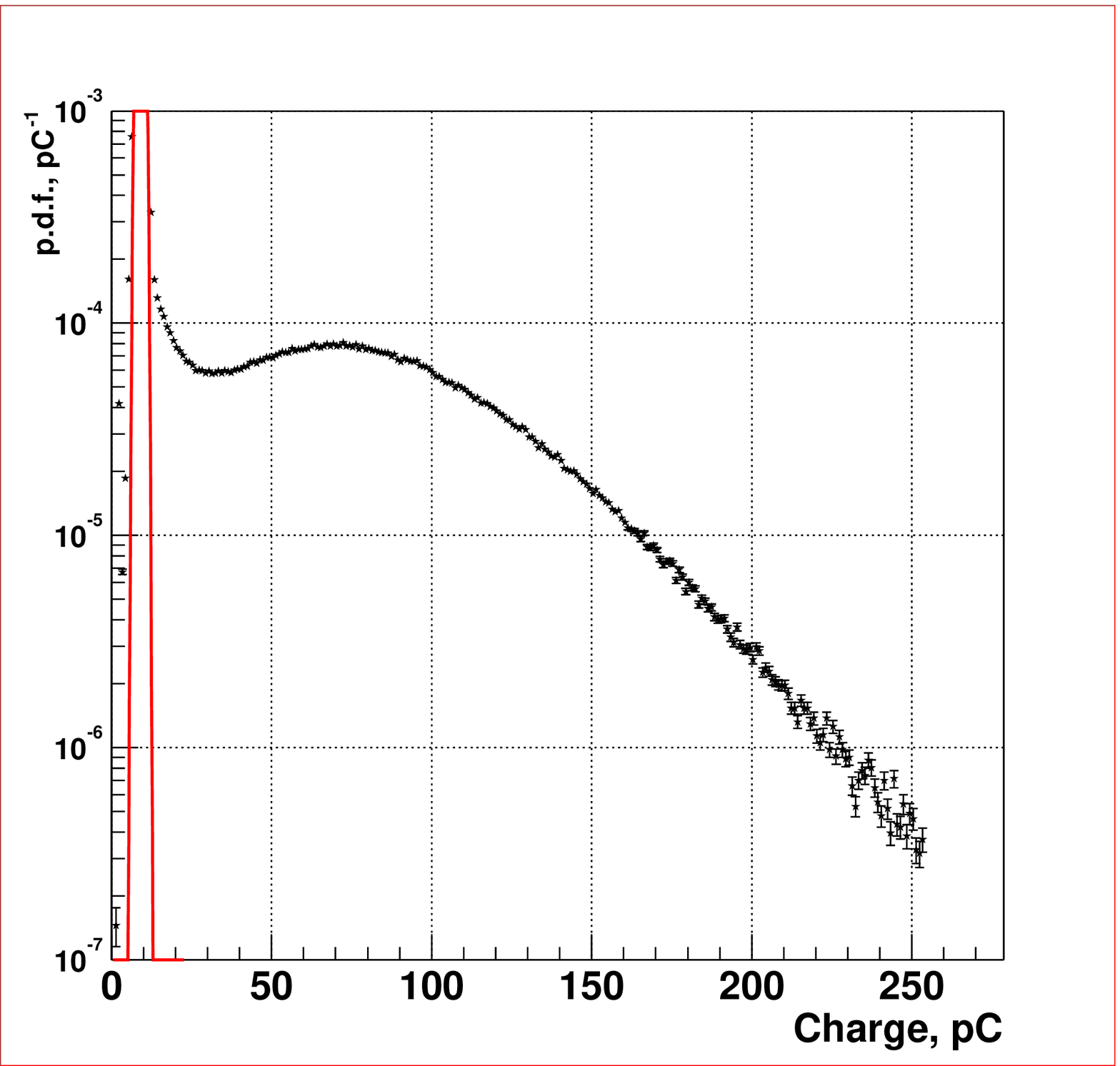}\end{center}

\caption{\label{Figure:Adc2795_bad}Example of the charge spectrum of the
PMT with a poor peak-to-valley ratio ($p/V=1.35$ and $v_{1}=0.3$).}
\end{figure*}

\begin{figure*}
\begin{center}\includegraphics[%
  width=0.80\textwidth,
  height=0.40\textwidth]{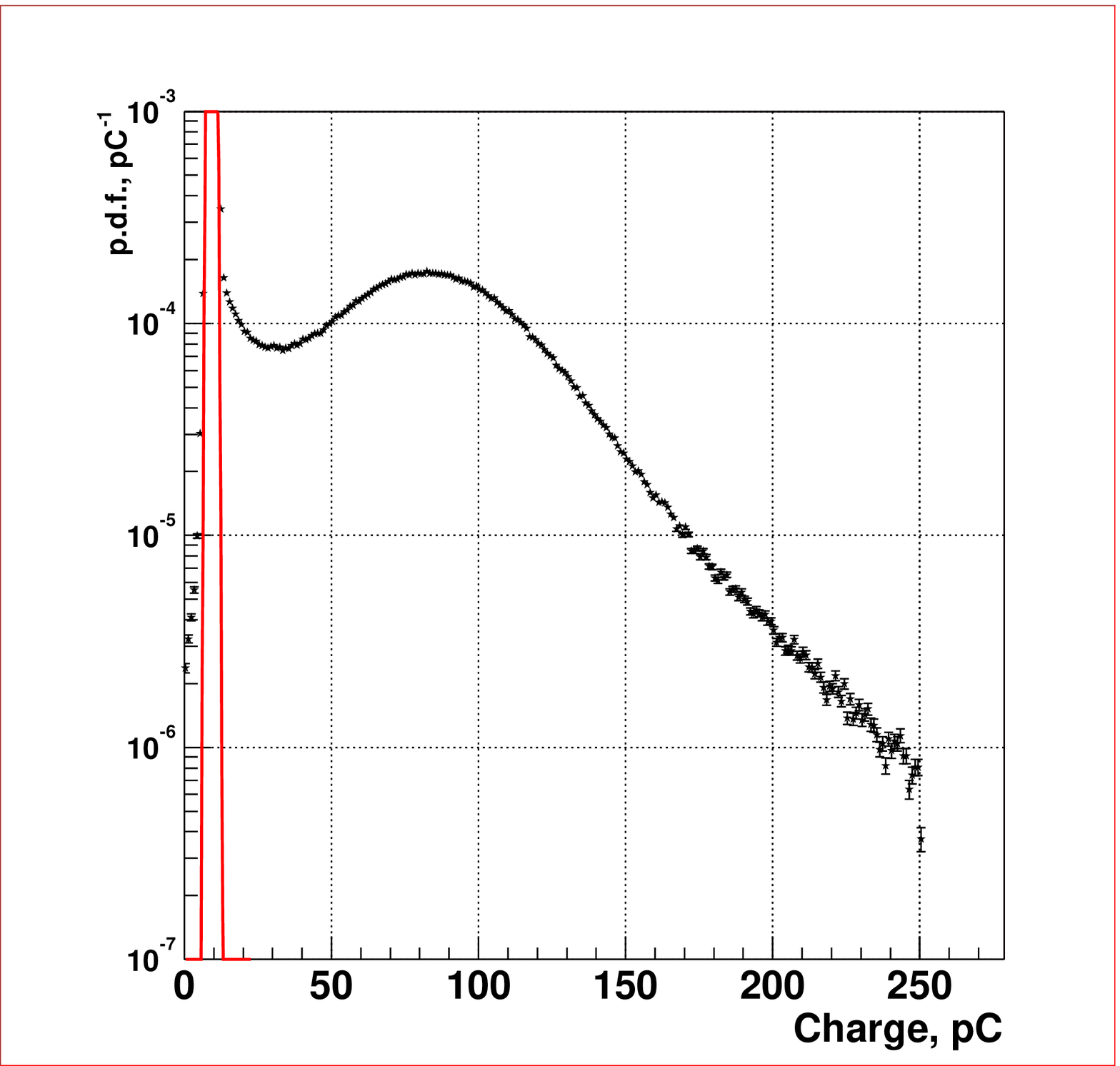}\end{center}

\caption{\label{Figure:Adc2885_good}Example of the charge spectrum of the
PMT with a high peak-to-valley ratio ($p/V=2.27$ and $v_{1}=0.46$).}
\end{figure*}

The charge spectra of the 8'' ETL9351 series were studied at our
test facility before the bulk test of the PMTs start. The results
were presented in \cite{Filters}, where a simple phenomenological
model was proposed to describe the data. In particular, it was shown
that the single photoelectron amplitude spectrum of the PMT consists
of two main contributions. The main contribution (>80\% of all signals)
is described with a Gaussian distribution; the remaining $\simeq20\%$
of signals are underamplified and can be described by an exponential
distribution with a negative slope. Examples of various charge spectra
are shown in Fig.\ref{Figure:Adc2795_bad} (PMT with a poor resolution)
and Fig.\ref{Figure:Adc2885_good} (PMT with a good resolution).

The data of the charge spectra for 2200 PMTs were averaged in order
to obtain a typical response of the photomultiplier. The procedure
of spectra averaging consisted of the following steps:

\begin{enumerate}
\item The position of the pedestal in the charge spectrum is found and the
histogram is shifted in order to put its pedestal at the position
corresponding to $q=0$. The histogram is normalized to 1.
\item All the histograms are summed together and normalized to 1 once more.
The obtained histogram contains the mean characteristics of a sample
of the PMTs used with a pedestal position at $q=0$.
\end{enumerate}
An ideal single electron charge response model consisting of a Gaussian
and an exponential has been used to fit the experimental data (\cite{Filters}):

\begin{flushleft}\begin{equation}
SER_{0}(q)=\left\{ \begin{array}{cc}
\frac{p_{E}}{A}e^{-\frac{q-q_{p}}{A}}+\frac{1}{\sqrt{2\pi}\sigma_{0}}\frac{1-p_{E}}{g_{N}}e^{-\frac{1}{2}(\frac{q-q_{0}-q_{p}}{\sigma_{0}})^{2}}, & q>q_{p}\\
0, & q\leq q_{p}\end{array}\right.\label{Eq:Ser0}\end{equation}
\end{flushleft}

with the following parameters:

- $A$ is the slope of the exponential part of the $SER_{0}(q)$

- $p_{E}$ is the fraction of events under the exponential function,

- $q_{p}$ is the pedestal position,

- $q_{0}$ and $\sigma_{0}$ the mean value and the standard deviation
of the Gaussian part of the single p.e. response respectively;

and the factor \[
g_{N}=\frac{1}{2}\Big(1+Erf\Big(\frac{q_{0}}{\sqrt{2}\sigma}\Big)\Big)\]
 takes account for the cut of the Gaussian part of the PMT response.

To account for the electronics noise one should perform the convolution
of the ideal SER with a noise function, $Noise(x)$ : \[
SER(q)=SER_{0}(q)\otimes Noise(q),\]
 where: \[
Noise(q)=\frac{1}{\sqrt{2\pi}\sigma_{p}}e^{-\frac{1}{2}(\frac{q-q_{p}}{\sigma_{p}})^{2}},\]
 which fits the pedestal with a proper normalization. The convolution
does not influence the Gaussian part of the SER since $\sigma_{1}>>\sigma_{p}$
(in our measurements $\sigma_{p}\sim0.01\sigma_{1}$), but it does
affect the exponential one which is closer to the pedestal. The analytical
formula for the convolution of the exponential function with the Gaussian
is: \[
\frac{p_{E}}{2A}\cdot e^{\frac{\sigma_{p}^{2}-2A(q-q_{p})}{2A^{2}}}\cdot(1+Erf(\frac{A(q-q_{p})-\sigma_{p}^{2}}{\sqrt{2}A\sigma_{p}})),\]
 where $Erf(q)$ is the error function.

The PMT response for a low light intensity contains a certain amount
of multiple primary p.e. signals. Assuming the linearity of the PMT
response, one can write: $q_{n}=nq_{1}$ and $\sigma_{n}=\sqrt{n}\sigma_{1}$,
where $q_{n}$ and $\sigma_{n}$ are the mean value and the standard
deviation of the PMT response to n p.e., respectively. Taking into
account the Poisson distribution of the detected light and using a
Gaussian approximation for the responses to $Np.e.>2$, the multi-p.e.
response will have the following form: \begin{equation}
M(q)=\sum_{n=2}^{N_{M}}\frac{P(n;\mu)}{\sqrt{2n\pi}\sigma_{1}}e^{-\frac{1}{2n}(\frac{q-nq_{1}-q_{p}}{\sigma_{1}})^{2}}\label{Formula:Mq}\end{equation}
 where the response to n p.e. is approximated by a Gaussian and $P(n;\mu)$
is the Poisson distribution with mean value $\mu$ to account for
the different contributions of $0\rightarrow n$ p.e. In (\ref{Formula:Mq})
$N_{M}$, the maximum number of multiple-p.e. responses considered,
depends on $\mu$ and on the ADC scale. The function $M(q)$ has three
additional parameters $\mu$, $q_{1}$ and $\sigma_{1}$.

The approximate values of $q_{1}$ and $\sigma_{1}$ can be calculated
from (\ref{Eq:Ser0}): \[
q_{1}\approx(1-p_{E})\cdot q_{0}+p_{E}A\]
 \[
\sigma_{1}^{2}\approx(1-p_{E})\cdot(\sigma_{0}^{2}+x_{0}^{2})+2p_{E}A^{2}-q_{1}^{2}.\]
 The approximate character of these formulae come from the cut in
the Gaussian part of the SER, whose portion below 0 is truncated.

The final fitting function for the PMT spectrum can be written as:
\begin{equation}
f(x)=N_{0}\cdot(P(0)\cdot Noise(x)+P(1)\cdot SER(x)+M(x))\label{FittingFunction}\end{equation}
 where $N_{0}$ is a normalization factor. 

\begin{figure*}
\begin{center}\includegraphics[%
  width=0.80\textwidth,
  height=0.40\textwidth]{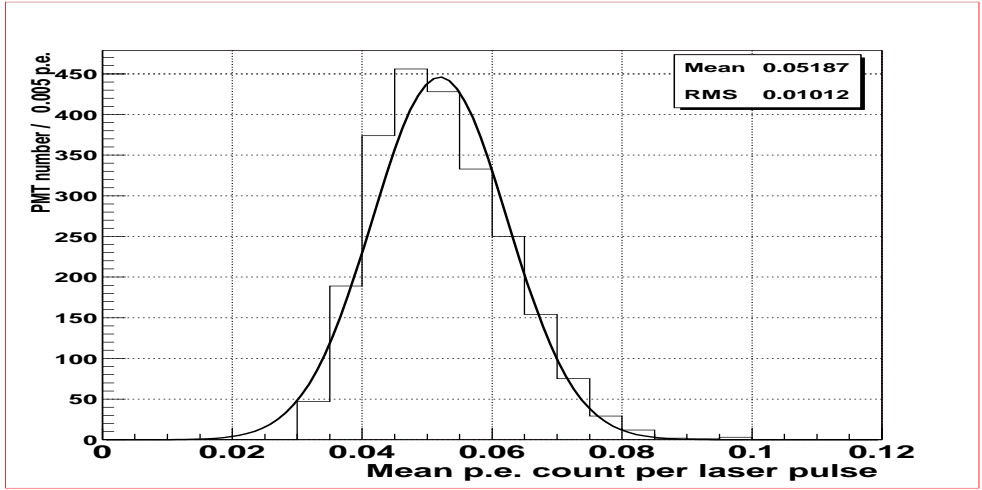}\end{center}

\caption{\label{Fig:MuStat}mean photoelectrons count per laser pulse. The
Gaussian fit is superimposed with the parameters practically coinciding
with the mean and r.m.s. value for the histogram.}
\end{figure*}

A certain distortion of the spectra is expected due to the different
PMTs illumination, sensitivity and amplification. All the electronics
channels were carefully calibrated in order to provide exact knowledge
of the amplification of the channel and the high voltage of the PMTs
were adjusted with a precision of better than $2\%$ \cite{TestFacility,HV}.
Thus, the main expected source of the spectrum distortions is due
to the effect of different PMT mean counts. The mean photoelectrons
count was checked during the tests with a high precision, the results
are presented in Fig.\ref{Fig:MuStat}. In order to account for the
distortions of the spectrum due to the variations of the mean photoelectron
count, the Poisson probabilities of formula (\ref{FittingFunction})
were transformed using the rule:

\[
P_{exp}(N)=\int_{0}^{N_{max}}P(N,\mu)F_{\mu}(\mu)d\mu,\]

with $N_{max}\simeq10$, which is big enough for the mean count of
$\mu=0.05$ p.e. and $F_{\mu}(\mu)$ is the measured distribution
of the mean photoelectrons count (see Fig.\ref{Fig:MuStat}). A Gaussian
distribution was used as an approximation of $F_{\mu}(\mu)$.

\begin{figure*}
\begin{center}\includegraphics[%
  width=1.0\textwidth,
  height=0.45\textwidth]{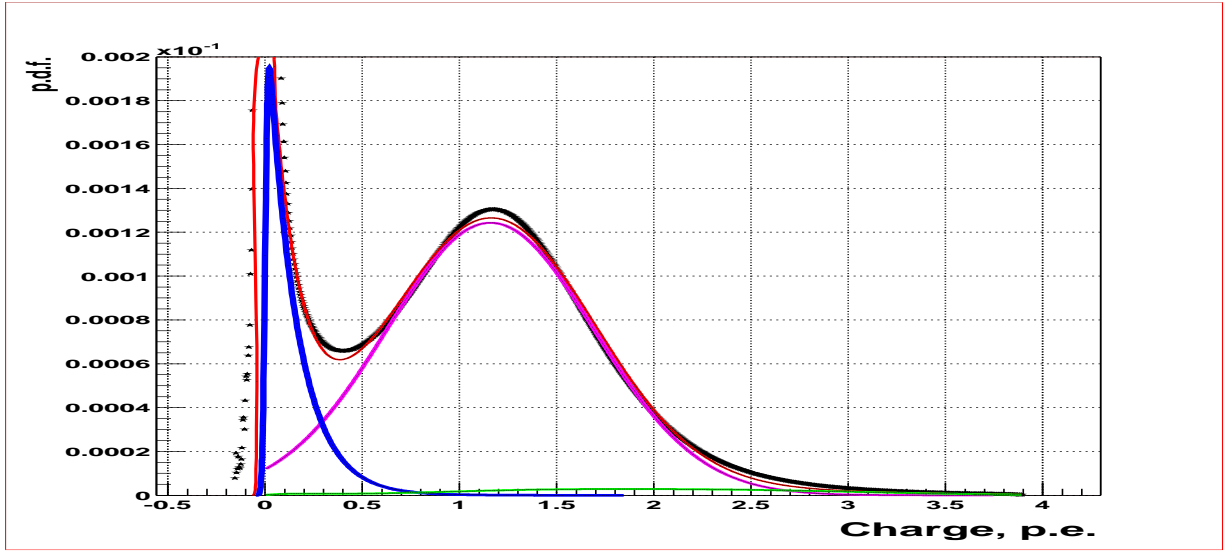}\end{center}

\caption{\label{Figure:ADC_fit}The fit of the average charge spectra of the
PMT.}
\end{figure*}

The results of the fit are presented in Fig.\ref{Figure:ADC_fit}
and the parameters are given in Table.\ref{Table:FitResults}. It
should be noted that parameters $q_{1}$ and $\sigma_{1}$ are not
independent. The value of $q_{1}$ was used to calibrate the scale
in p.e., thus $q_{1}=1.00$ by definition. 

\begin{table*}
\begin{center}\begin{tabular}{|c|c|c|c|c|c|c|c|}
\hline 
parameter&
$q_{0}$&
$\sigma_{0}$&
A&
$p_{U}$&
$\mu$&
$q_{1}$&
$\sigma_{1}$\tabularnewline
\hline
value&
1.16 p.e.&
0.52 p.e.&
0.19 p.e.&
0.18&
0.059 p.e.&
1.00 p.e.&
0.58p.e.\tabularnewline
\hline
\end{tabular}\end{center}

\caption{\label{Table:FitResults}The parameters of the charge spectrum fit}
\end{table*}

The mechanism of the electron multiplication allows to extract the
gain on the first PMT dynode $g_{1}$. First, we note that the variance
of the normally multiplied PMT signal (i.e. the part of the charge
spectrum corresponding to a Gaussian contribute for which $v(M)\equiv v_{0}=(\frac{\sigma_{0}}{q_{0}})^{2}$)
should satisfy the relation for the cascade processes (see i.e.\cite{birks}):

\begin{equation}
v(M)=v(g_{1})+\frac{v(g_{2})}{g_{1}}+\frac{v(g_{3})}{g_{1}g_{2}}+...,\label{Formula_vM}\end{equation}

where $M$ is total multiplication of the 12-stages dynode system,
$g_{1}$ is the mean electron multiplication factor on the first dynode,
$g_{2}$- on the second, etc. The divider used provided equal voltage
on the last 9 stages of multiplication $U_{D}$, double voltage on
the second stage of multiplication, and $1.5U_{D}$ on the third one.
If $U_{D}$ is small enough to ensure the linearity of the multiplication
in the region up to $2U_{D}$, then $g_{2}=2\cdot g$ and $g_{3}=1.5\cdot g$,
where $g$ is the mean amplification on each of the remaining dynodes.
In the case of the Poisson variances of the electron multiplication
$v(g_{1})=\frac{1}{g_{1}}$, $v(g_{2})=\frac{1}{g_{2}}$ etc., and
(\ref{Formula_vM}) reduces to:

\begin{equation}
v_{0}\equiv(\frac{\sigma_{0}}{q_{0}})^{2}=\frac{1}{g_{1}}\cdot\frac{6g^{2}-3g-1}{6g(g-1)}.\label{Formula_vM2}\end{equation}

\begin{figure*}
\begin{center}\includegraphics[%
  width=0.80\textwidth,
  height=0.40\textwidth]{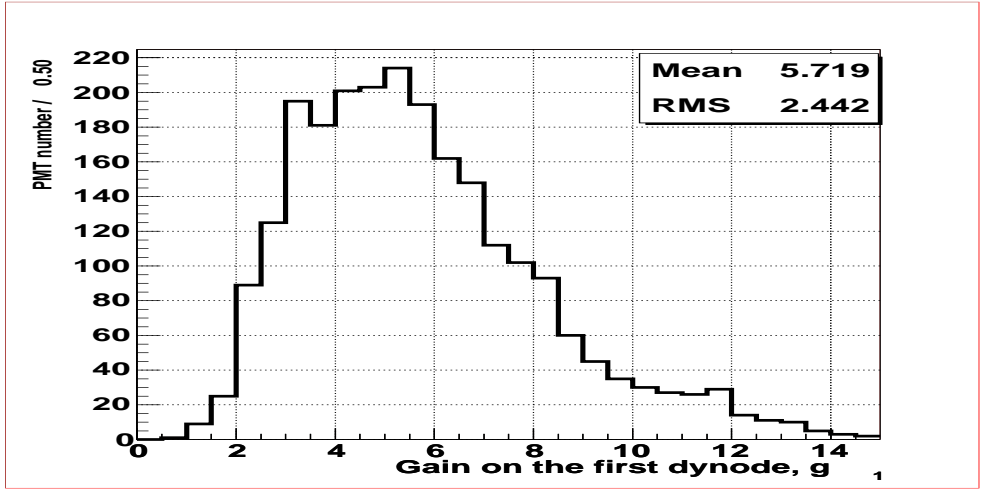}\end{center}

\caption{\label{Figure:g1}The statistics of the mean electrons gain on the
first dynode, $g_{1}$}
\end{figure*}

\begin{figure*}
\begin{center}\includegraphics[%
  width=0.80\textwidth,
  height=0.40\textwidth]{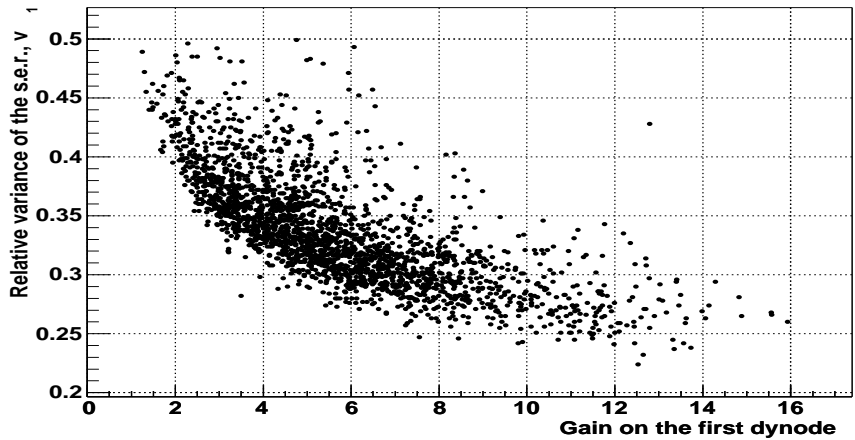}\end{center}

\caption{\label{Figure:v1_vs_g1}Relative variance of the single electron
charge spectrum in dependence on the gain on the first dynode.}
\end{figure*}

The mean gain of the multiplier is set to $M=2\times10^{7}$, and
the overall gain of the multiplier is the product of the gains at
each stage: $M=3g_{1}g^{11}$ (on the last stages of the electron
multiplier the gain can be reduced because of the spatial charge effect,
as it was noted in \cite{HV}, but it is of no significance for our
calculations), and we have a system of two equations with two variables.
Because of the 11-th degree in the equation for $g$ it is practically
independent of $g_{1}$and lies in the region $3.6<g<3.3$ for $5<g_{1}<15$.
The estimation of $g_{1}$ for 2000 PMTs is presented in Fig.\ref{Figure:g1},
in Fig.\ref{Figure:v1_vs_g1} is presented the relative variance of
the single electron charge spectrum in dependence on the gain on the
first dynode. One can see, that the overall performance of the PMT
depends on the gain on the first dynode. The average value of $v_{1}=0.34$
corresponds to the quite low mean values of $g_{1}=4-6$. This value
is much less than expected for the type of the material used at the
high voltage applied. Probably, this is the effect of the different
conditions during the multiplication coefficient measurements, in
which the incidence angle of the primary electrons is fixed, and the
operational conditions in the PMT, when the angle of incidence is
varying in a wide interval, and the effective gain is a result of
averaging over all possible angles of incidence. Using the value $g=3.56$
(for $g_{1}=5.7$) one can write an approximate formula for the estimation
of $g_{1}$ from the relative width of the main peak at the s.e.r.
spectrum: $g_{1}\simeq\frac{1.18}{v_{0}}$. 

The relatively big amount of the small amplitude pulses is due to
the fact that all the events in the 60 ns window are processed. As
it will be discussed below, there is a notable amount of small amplitude
pulses which arrives with a delay in respect to the main pulse. They
correspond to the inelastic scattering of photoelectron on the first
dynode. A special measurement was performed with a selection on the
arrival time, showing the decrease of the contribution of the small
amplitude pulses for the event arriving in time interval $t_{0}\pm\sigma_{t}$.
The advanced study of the time-amplitude characteristics of the PMTs
is now in progress, the results will be presented elsewhere. 

It should be noted that the model fails to describe the s.e.r. in
all details. Though the response to the more intense light sources
with $\mu\simeq1$ is fitted well \cite{Filters}, the very high statistics
of the data allows to see the deviations from the model. One can note
three features of the model in comparison to the experimental data:
1) the underfilled valley region between the main peak and the pedestal;
2) the lower peak value of the model; 3) the tail of the model is
almost a factor 2 lower than the experimental data. The last feature
is a consequence of the Poisson statistics of the multiplication on
the first dynode. If one takes into consideration the mean value of
$g_{1}\simeq5.7$, then it is easy to check that at the position corresponding
to $>2\times g_{1}$ primary electrons the Poisson distribution has
factor 2 higher tail. The mean gain for the underamplified branch
should correspond to the value $1/g_{1}$ in the case when a primary
electron loses all its energy on the first dynode in inelastic scattering.
One can see that the parameter $A=0.19$ p.e. from Table \ref{Table:FitResults}
is very close to the value $0.2$, that can be obtained dividing the
mean amplification for a Gaussian part of the s.e.r. spectrum $q_{0}=1.16$
by the mean value of the $g_{1}$ coefficient $g_{1}=5.7$ (see below).
This fact confirms that the major part of the underamplified signals
are produced in totally inelastic scattering of the photoelectron
on the first dynode.

\subsection{Charge spectrum characterization}

\begin{figure*}
\begin{center}\includegraphics[%
  width=0.80\textwidth,
  height=0.35\textwidth]{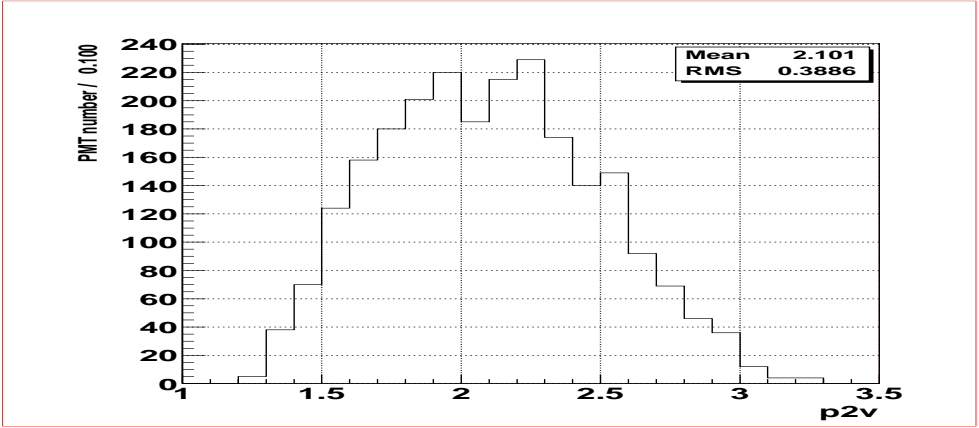}\end{center}

\caption{\label{Figure:p2v}Peak-to-valley ratio}
\end{figure*}

The PMT charge resolution is characterized by the manufacturer by
the peak-to-valley ratio. The measured peak-to-valley ratio is presented
in Fig.\ref{Figure:p2v}. %

\begin{figure*}
\begin{center}\includegraphics[%
  width=0.80\textwidth,
  height=0.35\textwidth]{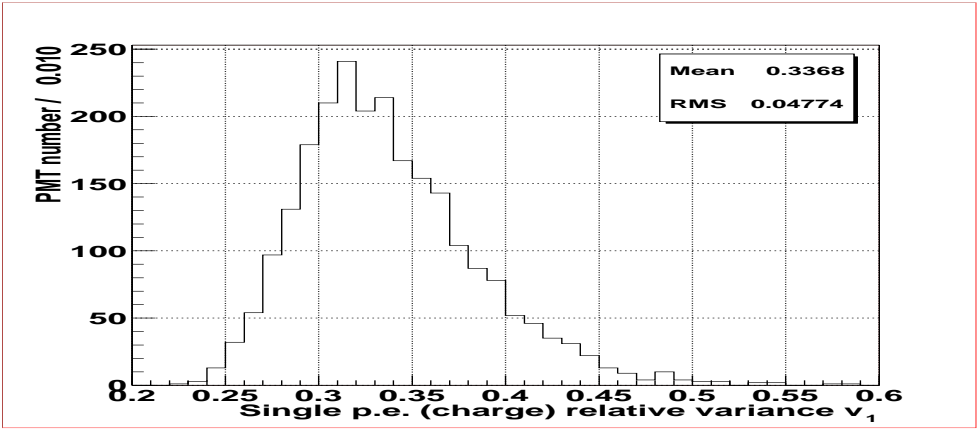}\end{center}

\caption{\label{Figure:v1}Relative variance of the single p.e. response}
\end{figure*}

\begin{figure*}
\begin{center}\includegraphics[%
  width=0.90\textwidth,
  height=0.45\textwidth]{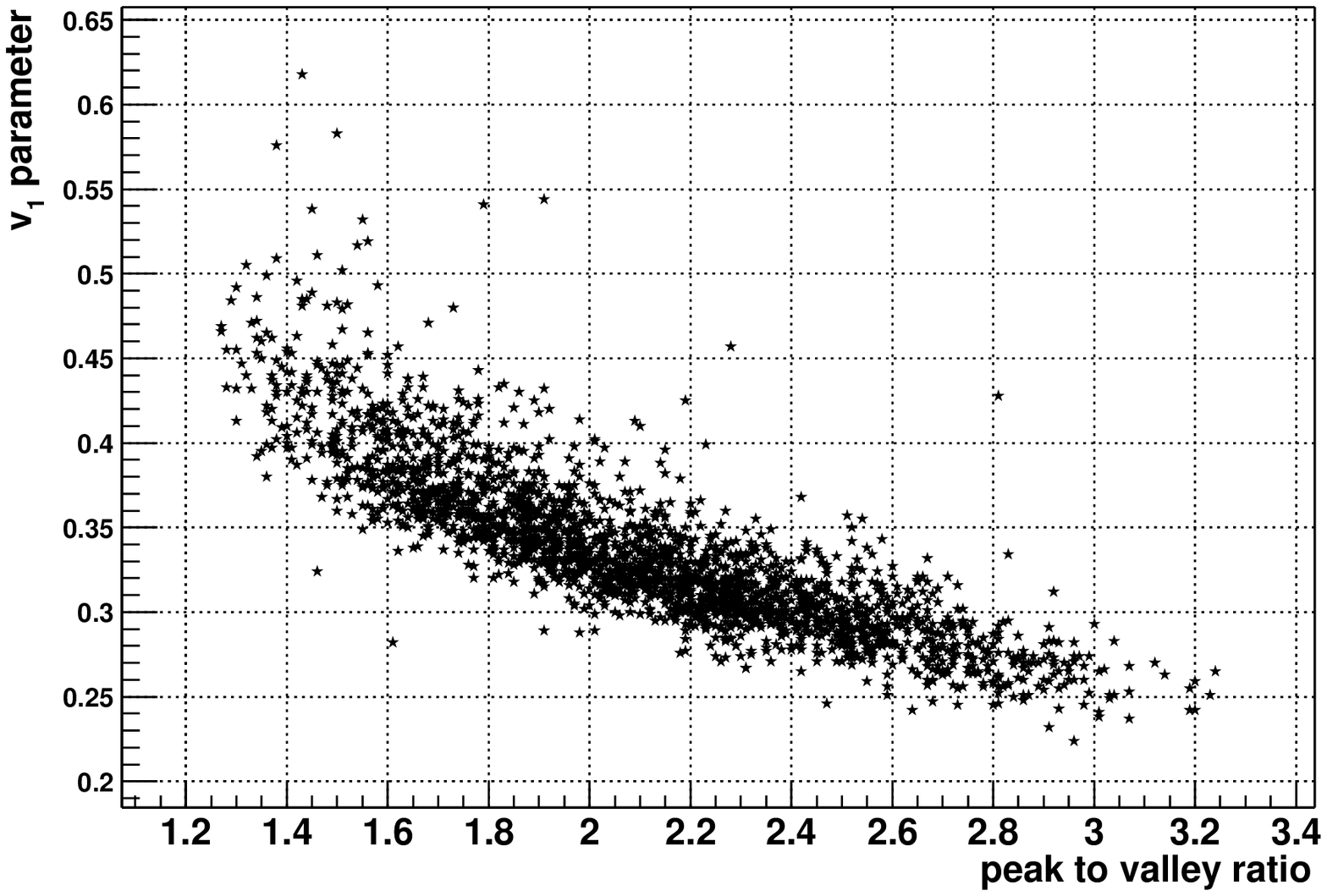}\end{center}

\caption{\label{Figure:v1vsp2v}Relative variance of the single p.e. response
versus peak to valley ratio.}
\end{figure*}

For the numerical estimation of the detector resolution this parameter
has no immediate significance. More informative is the relative resolution
of the single photoelectron charge spectrum $v_{1}\equiv\frac{\sigma_{1}^{2}}{q_{1}^{2}}$.
As it was shown in \cite{Resolutions} the relative charge resolution
of the detector with a spherical symmetry in the regime of the total
charge measurement can be related to the average relative resolution
of the single photoelectron charge spectrum $\overline{v_{1}}\approx\frac{1}{N_{PM}}\sum v_{1_{i}}$:

\textbf{\begin{equation}
R=\sqrt{\frac{\sigma_{Q}^{2}}{Q^{2}}}=\sqrt{\frac{1+\overline{v_{1}}}{Q_{0}(E)<f_{s}>_{V}}+v(f_{s})}\:.\label{R_V}\end{equation}
}with constant parameters $v(f_{s})$ and $<f_{s}>_{V}\simeq1$ characterizing
the detector. The parameters $v_{1}$ and peak-to-valley ratio are
correlated, as it can be seen in Fig.\ref{Figure:v1vsp2v}, where
the relative variance of the single p.e. response is plotted versus
the peak-to-valley ratio.

\begin{figure*}
\begin{center}\includegraphics[%
  width=0.80\textwidth,
  height=0.40\textwidth]{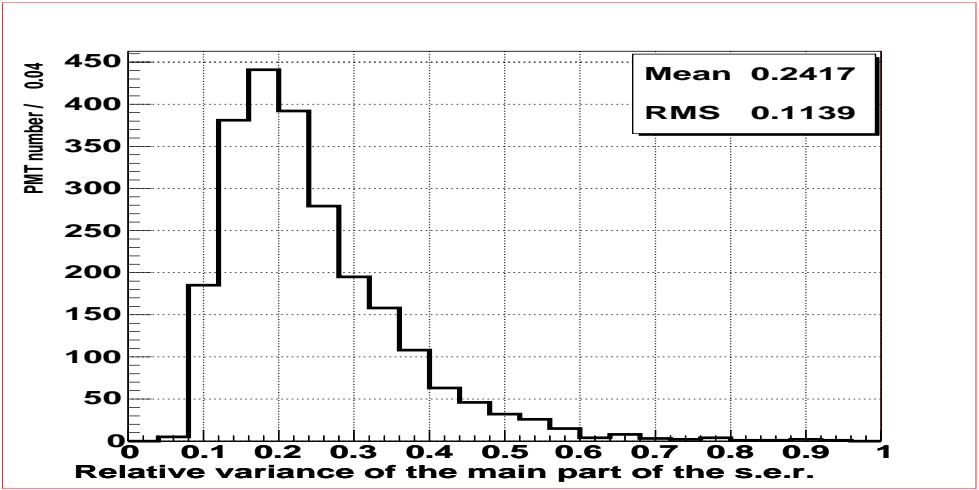}\end{center}

\caption{\label{Figure:v0}The mean relative variance of the Gaussian part
of the s.e.r.}
\end{figure*}

\begin{figure*}
\begin{center}\includegraphics[%
  width=0.80\textwidth,
  height=0.40\textwidth]{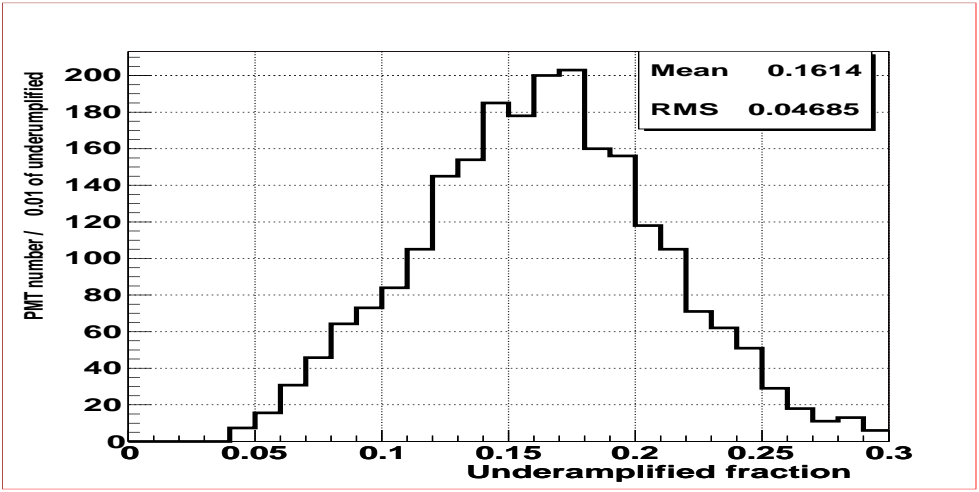}\end{center}

\caption{\label{Figure:pU}The fraction of the exponential branch}
\end{figure*}

The statistics of the measured values of parameter $v_{1}$ is presented
in Fig.\ref{Figure:v1}. The relative single photoelectron charge
variance averaged over all accepted PMTs turns out to be $\overline{v_{1}}\approx0.34$.
The mean relative variance of the Gaussian part of the s.e.r. is much
less, $\overline{v_{0}}=0.24$ (see Fig.\ref{Figure:v0}). The presence
of the exponential branch in the s.e.r. is the reason of such a resolution
degradation. The relative strength of the exponential branch (underamplified
signals fraction) is presented in Fig.\ref{Figure:pU}.

\section{Transit time}

\subsection{Time spectrum structure}

The most important characteristics of the time spectrum are the variance
of the main peak in the spectrum, the relative amount of the late
pulses and the relative amount of the prepulses.

The details of the transit time spectrum of the 8'' ETL9351 series
were studied in \cite{TDC} using the results of the bulk tests. The
spectra of 2200 PMTs were used to produce an averaged time spectrum.
Because every PMT operates at its own voltage, and the lighting conditions
depends on the position on the test tables, the procedure of averaging
should be preceded by equalizing the difference in the measurement
conditions. In \cite{TDC} it was shown that \foreignlanguage{british}{}the
probability density function of the single photoelectron time arrival
can be calculated as:

\begin{equation}
N_{1}(i)=\frac{1}{\mu}\frac{N_{Exp}(i)}{1-s\;(i)},\label{Formula:Hist}\end{equation}

where \begin{equation}
s\:(i)\equiv\frac{1}{N_{Triggers}}\sum_{k=1}^{k=i}N_{Exp}(k)\label{Formula:RunningSum}\end{equation}
is the running sum of the histograms of the experimental data $N_{Exp}(i)$
normalized by the number of the system starts $N_{Triggers}$. For
$N_{Triggers}$ large enough $s\:(\infty)=1-e^{-\mu}$. 

\begin{figure*}
\begin{center}\includegraphics[%
  width=1.0\textwidth,
  height=0.45\textwidth]{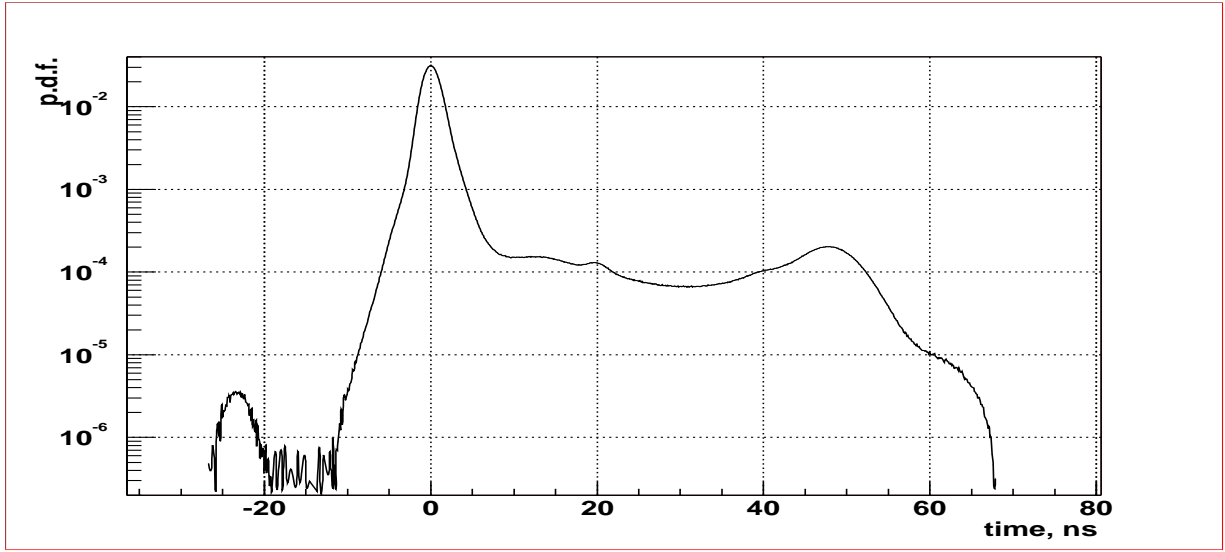}\end{center}

\caption{\label{Figure:TDC}The averaged timing characteristics of the ETL9351
PMT. }
\end{figure*}

The procedure of spectra averaging consisted in the following steps:

\begin{enumerate}
\item Using the measured value of the dark rate, the contribution $N_{dark}$
of the random coincidences at one bin was calculated.
\item Using equations (\ref{Formula:Hist}) and (\ref{Formula:RunningSum})
with $N_{exp}(i)$ substituted by $N_{exp}(i)$-$N_{dark}(i)$, the
$N_{1}(i)$ function was calculated and normalized: \begin{equation}
n_{1}=\frac{N_{1}(i)}{\sum_{i=1}^{N_{bins}}N_{1}(i)}.\label{Formula:n1}\end{equation}

\item The peak in the distribution $n_{1}$ is found and the histogram is
shifted in order to put its maximum at the position corresponding
to $t=0$. 
\item All the histograms are summed together and normalized to 1 once more.
The obtained histogram contains the mean characteristics of a sample
of the PMTs used with a peak (not mean time of the arrival) at the
position $t=0$.
\end{enumerate}
The resulting histogram is presented in Fig.\ref{Figure:TDC}. This
is the PMT transit time p.d.f. averaged over a 2200 PMT sample. The
spectrum corresponds to an average level of the discriminator of $0.17$
p.e. The structure of the photomultiplier transit time exhibits the
following main features: 1) an almost Gaussian peak at the position
$t=0$ ns; 2) a very weak peak at $t=-24$ ns; 3) a weak peak at $t=48$
ns; 4) the continuous distribution of the signals arriving between
the main peak and the peak at $t=48$ ns; and 5) another very weak
peak at $t=20$ ns. 

Below we summarize briefly the results of the study of these features
performed in \cite{TDC}. A small peak at about -24 ns is a result
of direct photoproduction of the electron on the first dynode. The
amplitude of these pulses is a factor $g_{1}$(amplification of the
first dynode) lower then the amplitude of the main peak. Because a
typical value is $g_{1}\simeq5.7$, these pulses are strongly suppressed
by the discriminator threshold set at the 0.2 p.e. level. The shape
of the peak is well approximated by a Gaussian one with a total intensity
of $1.22\cdot10^{-4}$. \.{ }This corresponds roughly to the ratio
of the surfaces of the photocathode $S_{PMT}$ and the first dynode
$S_{D_{1}}$. One can write the relative probability of the photoelectron
production at the first dynode assuming the same efficiency of the
photoproduction at the photocathode and the first dynode as:

\[
p=\frac{S_{D_{1}}}{S_{PMT}}\cdot\tau\cdot r\cdot e^{-\frac{th}{1/g_{1}}},\]

where $\tau$ is transparency of the photocathode; $r$ is the ratio
of the quantum efficiencies on the first dynode and on the photocathode,
and $th$ is the threshold of the detection (in assumption of the
exponential distribution of the amplitudes of prepulses). An estimation
with $\tau=0.5$, $th=0.17$ p.e., $r=1$ and $g_{1}=5.7$ gives $p\simeq1.5\cdot10^{-4}$
which is in agreement with the measured value.

The difference $dt=23.2$ ns between the position of the main peak
$t_{0}$, and the position of the prepulses peak $t_{pp}$ corresponds
to the drift time of the electron from the photocathode to the first
dynode $t_{d}$ with the time of flight of photon to the first dynode
$t_{tof}$ subtracted: $dt=t_{0}-t_{pp}=t_{d}-t_{tof}$$\,$. The
time of flight can be calculated from the known distance between the
photocathode and the first dynode, which is 123 mm (radius of the
spherical photocathode is 110 mm, the focusing grid is situated at
the center of the sphere, the distance between the focusing grid and
the first dynode is 13 mm). Hence the time of flight of photon inside
the PMT is $tof=0.4$ ns, and the drift time $t_{d}=dt+tof=23.6$
ns. The drift time is the same for all the PMTs tested, because the
potential difference between the photocathode and the first dynode
is stabilized.

In setups with a large number of PMTs in use, such as the Borexino
detector, the presence of prepulses is a potential source of early
triggers of the system. With a total number of the 2000 PMTs and the
light yield of 400 p.e./MeV, the probability of early trigger for
an event of 500 keV is about $2.5\%$.

The peak with probability $p_{r}=5.8\cdot10^{-4}$ at $t=20.2$ ns
corresponds to the electron's transit time from the photocathode to
the focusing grid and is caused by light feedback on the accelerating
dynode. The spread of the peak $\sigma=1.07$ ns coincides with the
main peak spread. 

Pulses arriving with a delay relative to the main pulse in the time
spectrum are called late pulses. The position of the last peak helps
in clarifying its origin. The difference between the position of the
last peak and the main peak is $\Delta t=47.6$ ns, and it coincides
with a double drift time $2t_{d}=47.2$ ns. The double drift time
can be explained by electrons which elastically backscatter on the
accelerating grid, then go away from it, slow down and stop near the
photocathode, and then are accelerated back to the first dynode to
produce a signal. Two experimental facts are confirming this conclusion.
First, the position of the last peak depends on the applied voltage,
in the case of the backscattering on the first dynode the drift time
should be constant because the potential of the first dynode is stabilized.
The second argument is the geometry of the PMT dynode system entrance.
The first dynode is designed to scatter electrons by 90$^{\circ}$
angle. The elastically scattered photoelectrons should be transferred
to the second dynode with a higher speed \cite{Candy}, producing
small amplitude pulses preceding the main one. This phenomena was
indeed observed in our measurements (see \cite{TDC}), and we called
it early pulsing. 

The amplitude spectrum of these pulses is very similar to that of
the main peak pulses. The total probability to observe photoelectron
elastically scattered on the accelerating electrode is $0.024$. 

The shape of the main pulse has significant deviations from the Gaussian
shape not only in the region $t<t_{0}$, but in the region $t>t_{0}$,
as well. The distortions of the shape close to the main peak position
are due to the registering of the photons scattered inside the PMT.
The typical delay time of this signals is of the order of 1 ns.

The remaining contribution to the late pulses corresponds to an inelastic
scattering of the photoelectron on the first dynode, without any secondaries
produced. In this case, part of initial energy of the incident electron
is dissipated as heat in the material of the dynode, and the drift
time of the electron in this case depends on the remaining part of
the energy, and, naturally, is less than in the case of elastic scattering.
In the extreme case all the energy is dissipated, and, without any
delay, the electron is transferred to the next stage of amplification,
producing on the average a signal with an amplitude of factor $g_{1}$
smaller than a normal signal. In the intermediate case, the scattered
electron is delayed by a time in the range $0-2t_{d}$, and after
returning back to the first dynode produces a signal with lower amplitude
in comparison to the amplitudes of the main peak signals. The smaller
the delay, the smaller is the amplitude of the signal. The total probability
of inelastic scattering of photoelectrons is $0.038$.

\subsection{Time spectrum Characterization}

\begin{figure*}
\begin{center}\includegraphics[%
  width=1.0\textwidth,
  height=0.45\textwidth]{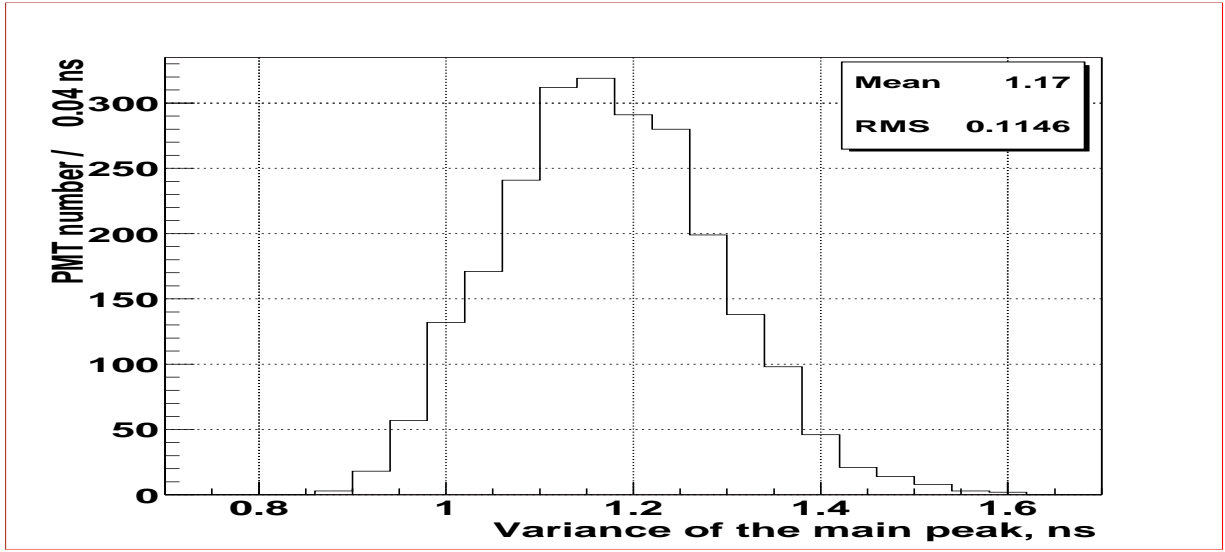}\end{center}

\caption{\label{Figure:MainPeakStat}Statistics of the main peak width}
\end{figure*}

The characteristic of main concern is the width of the main peak in
the transit time spectra. The results of measurements are presented
in Fig.\ref{Figure:MainPeakStat}. The width of the peak was defined
by fitting the main peak region with a Gaussian distribution.

As it was noticed above, the prepulses corresponding to the direct
photoproduction on the first dynode have a negligible probability
due to the small size of the first dynode in comparison to the photocathode
area. This pulses have small amplitude and can easily be discriminated.

\begin{figure*}
\begin{center}\includegraphics[%
  width=1.0\textwidth,
  height=0.45\textwidth]{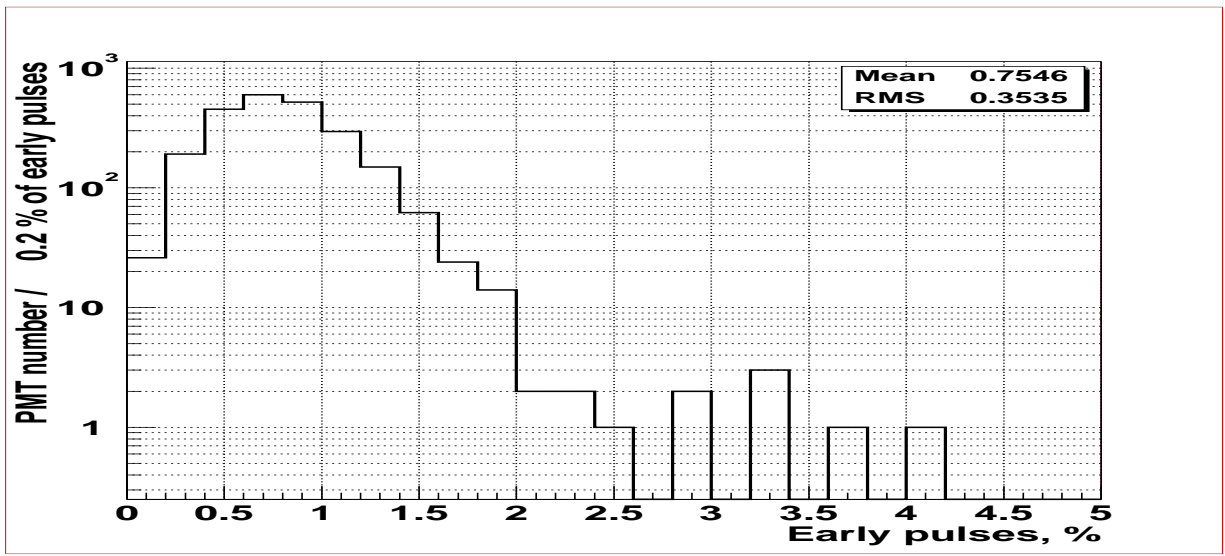}\end{center}

\caption{\label{Figure:EarlyPulsesStat}Statistics of the early pulses probability}
\end{figure*}

The more important characteristic in our case is the amount of the
early pulses, preceding the main peak in the time spectrum by a few
nanoseconds. We characterize their amount by the percentage of pulses
in the interval up to $t_{0}-3\sigma$, where $t_{0}$ is the position
of the main peak and $\sigma$ is the width of the main peak. The
results are shown in Fig.\ref{Figure:EarlyPulsesStat}. 

\begin{figure*}
\begin{center}\includegraphics[%
  width=1.0\textwidth,
  height=0.45\textwidth]{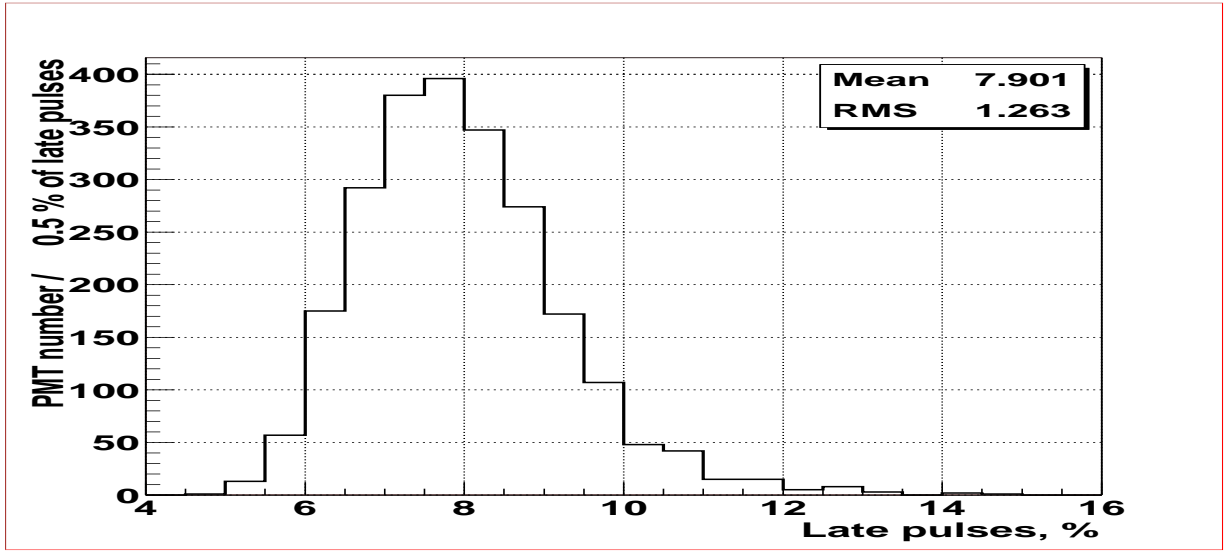}\end{center}

\caption{\label{Figure:LateStat}Statistics of the late pulses probability}
\end{figure*}

And, finally, the amount of the late pulses, arriving with a delay
greater than $3\sigma$ with respect to the main peak in the time
spectrum, is the last characteristic of the time spectrum of PMT.
The results are shown in Fig.\ref{Figure:LateStat}. The mean observed
amount of late pulses is $7.9\%$ and corresponds to the CFD threshold
set at the level of 0.17 p.e. The total probability to observe elastically
scattered photoelectron, as follows from the analysis of the average
transit time spectra, is $2.4\%$. This value is much closer to the
one measured in ETL factory tests. The reasons for this discrepancy
is the lower threshold in our measurements. The direct check with
a CFD threshold set at the level of 0.3.-0.4 p.e. gives much lower
values for the late pulses.

\section{Afterpulses}

\subsection{Afterpulses spectrum structure}

A photoelectron striking a residual gas trapped in the photomultiplier
can ionize a gas molecule, the ionized molecule will be accelerated
back to the cathode releasing several photoelectrons, thus forming
an undesirable satellite pulse \cite{Coates-Afterpulses}. The study
of the ionic afterpulses was performed e.g. in \cite{Torre}. In \cite{Afterpulses1}
afterpulse formation have been systematically studied directly introducing
the trace amounts of various gases into photomultiplier tubes. A photomultiplier
exhibit the phenomenon of afterpulses in the microsecond range with
total afterpulses rate $p_{a}$ ranging from 0.1\% to 5\%. The upper
limit usually is guaranteed by the manufacturer for the specified
HV. It should be noted, that the origin of the afterpulses in PMTs
is still being discussed.

The afterpulse rate $p_{a}$ is usually cited for the single photoelectron
response and a certain time region after the main pulse. If a PMT
is hit by a light pulse producing precisely \textbf{n} p.e. and the
afterpulse production is independent for each initial electron, then
the mean number of afterpulses will be defined by the following relation:

\begin{equation}
p_{a}(n)=n\cdot p_{a}\label{Formula:AfterpulsesProb}\end{equation}

The value of $p_{a}(n)$ in (\ref{Formula:AfterpulsesProb}) in principle
can be greater than 1 for large $n$, but the value of $p_{a}$ can
not exceed 1, otherwise the PMT will go into autogenerating regime.

For the series of the PMT responses to the light source with a mean
intensity corresponding to $\mu$ p.e. with a Poisson law of the registered
p.e. one should provide proper averaging over the Poisson probabilities
of the registering precisely n p.e.:

\begin{equation}
p_{a}(\mu)=\frac{\sum P(n)p_{a}(n)}{1-P(0)}=\frac{\mu\cdot p_{a}}{1-e^{-\mu}}.\label{Formula:PoissonAfterpulsesProb}\end{equation}

The delimiter in (\ref{Formula:PoissonAfterpulsesProb}) reflects
the fact that the afterpulses by their definition are following the
main pulse. For the values $\mu\ll1$ (\ref{Formula:PoissonAfterpulsesProb})
will be reduced to: \begin{equation}
p_{a}(\mu)\simeq p_{a}.\label{Formula:paLowLight}\end{equation}

For flashes with a large number of initial photoelectrons secondary
afterpulses are probable. The probability of observing secondary pulse
is $p_{a}^{sec}=<N>\cdot p_{a}^{2}$, where $<N>$ is the mean number
of secondary electrons produced in the afterpulses event%
\footnote{It is reported in literature, that an ion hitting photocathode can
produce in average 3-4 p.e. \cite{Bartlett} %
}

Formulae (\ref{Formula:AfterpulsesProb})-(\ref{Formula:PoissonAfterpulsesProb})
are true as well for the afterpulses rate $p_{a}(\mu,t)\Delta t$
in the interval $\Delta t$ at any time $t$ under the same assumptions.
The last formula is the basic in the practical afterpulses measurements.
One can see that the afterpulse rate due to the low intensity light
source coincide with the rate of the afterpulses of the single p.e.
response. %
\begin{figure*}
\begin{center}\includegraphics[%
  width=1.0\textwidth,
  height=0.22\textwidth]{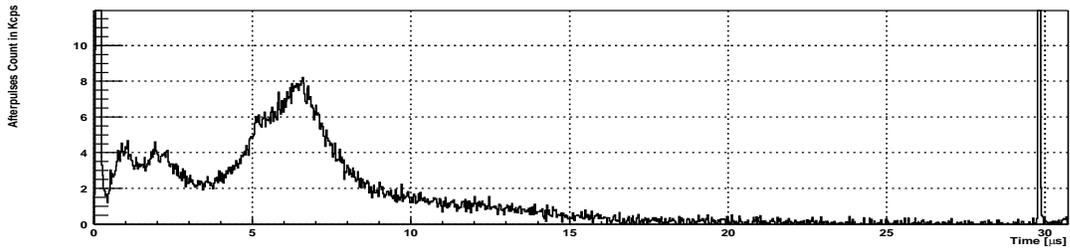}\end{center}

\begin{center}\includegraphics[%
  width=1.0\textwidth,
  height=0.22\textwidth]{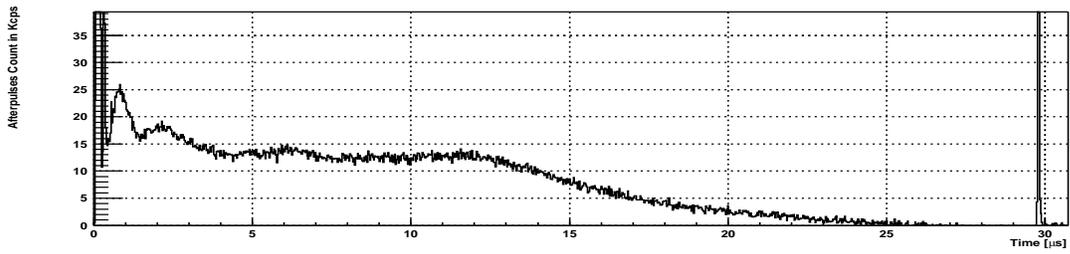}\end{center}

\begin{center}\includegraphics[%
  width=1.0\textwidth,
  height=0.22\textwidth]{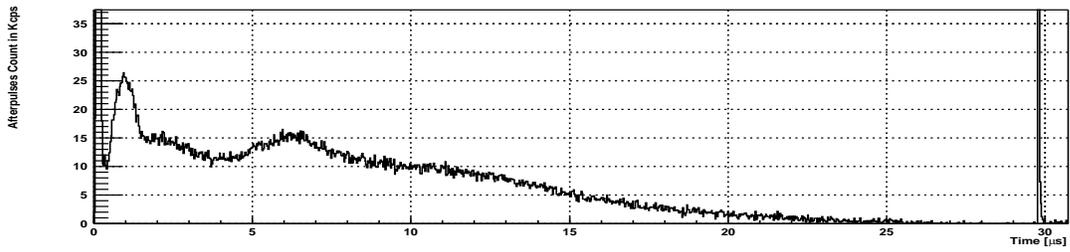}\end{center}

\begin{center}\includegraphics[%
  width=1.0\textwidth,
  height=0.22\textwidth]{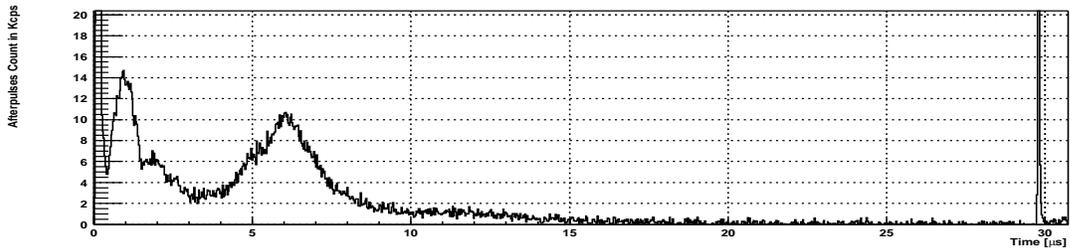}\end{center}

\caption{\label{Figure:Afterpulses}Measured afterpulse rate for 4 different
PMTs.}
\end{figure*}
 \foreignlanguage{english}{}

The afterpulse probability for the more than 2000 ETL PMTs has been
measured at the PMT test facility at the Gran Sasso Laboratory in
the frame of the Borexino program. It should be noted that Multihit
TDC was used in the measurements in contrast to the more common correlation
measurements with ordinary start-stop TDC or multichannel analyzers.
In such a way the acquired afterpulse spectra can in principle contain
multiple afterpulses (up to 15; the system was adjusted to measure
the time of arrival of the main pulse as the first one). The discriminator
threshold was set approximately at the level of 0.2 p.e. and was checked
out after the measurements. The mean measured threshold was $0.17$
p.e.

The afterpulse spectra for 4 different PMTs are presented in Fig.\ref{Figure:Afterpulses}.
The afterpulse rates are presented in the figure as the excess dark
rate counted in Kcps. The peak at $t\simeq30$ $\mu$s is due to the
next laser flash inside the MTDC gate (laser flashed at $\simeq33$
kHz). The amount of pulses in this peak has been used to check the
laser intensity. 

\begin{figure*}
\begin{center}\includegraphics[%
  width=1.0\textwidth,
  height=0.45\textwidth]{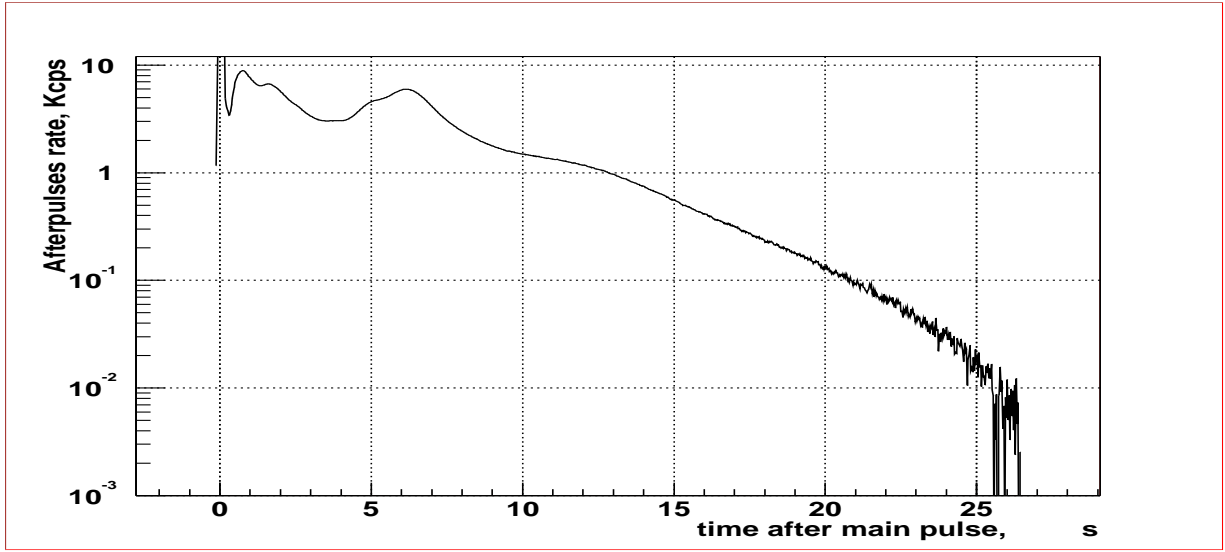}\end{center}

\caption{\label{Fig:AverageAP}Average spectrum of afterpulses.}
\end{figure*}

The spectra in Fig.\ref{Figure:Afterpulses} have common features:
relatively narrow peaks about 1 $\mu s$ and 2 $\mu$s, wider peaks
around 6.5 $\mu s$ and 12 $\mu s$ , and at $t>25$ $\mu s$ no afterpulses
are present. The analysis performed for the set of 2300 PMTs showed
no significant dependence of the afterpulses amount on the applied
voltage. The difference in the position of the peaks at 6 $\mu$s
for the PMT with working voltage 1200 V and 1800 V is only 150 ns.
This feature allows to perform averaging of the afterpulse spectra
over the set of all PMTs. The procedure of averaging is performed
as follows:

\begin{enumerate}
\item Using the measured value of the dark rate, the contribution $N_{dark}$
of the random coincidences at one bin was calculated and extracted
from the value at each bin.
\item For each histogram the position of the main peak in the distribution
and its integral are found and the histogram is shifted in order to
put its maximum at the position corresponding to $t=0$. The spectrum
is normalized on the integral of the main peak (number of true triggers).
\item All the histograms are summed together and normalized on the integral
of the main peak once more. The obtained histogram contains the mean
characteristics of a sample of the PMTs used with a main peak at the
position $t=0$.
\end{enumerate}
The resulting histogram is presented in Fig.\ref{Fig:AverageAP}.
The weak peak at 5 $\mu$s appears at the histogram together with
the already mentioned ones. The afterpulse amount is presented in
the histogram as an excess rate in Kcps.

The systematic study of the origin of the afterpulses is now in progress
and will be the subject of another article. Below we will give the
main afterpulse characteristics.

\subsection{Afterpulse characterization}

\begin{figure*}
\begin{center}\includegraphics[%
  width=0.60\textwidth,
  height=0.30\textwidth]{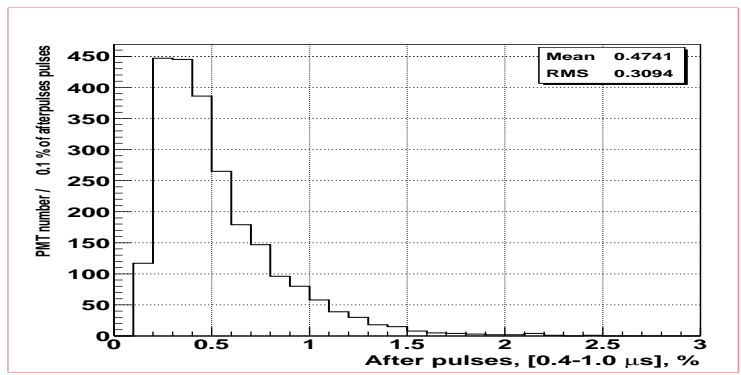}\end{center}

\caption{\label{Fig:ap1}Afterpulses in the region $0.4-1.0$ $\mu$s,}
\end{figure*}

\begin{figure*}
\begin{center}\includegraphics[%
  width=0.60\textwidth,
  height=0.30\textwidth]{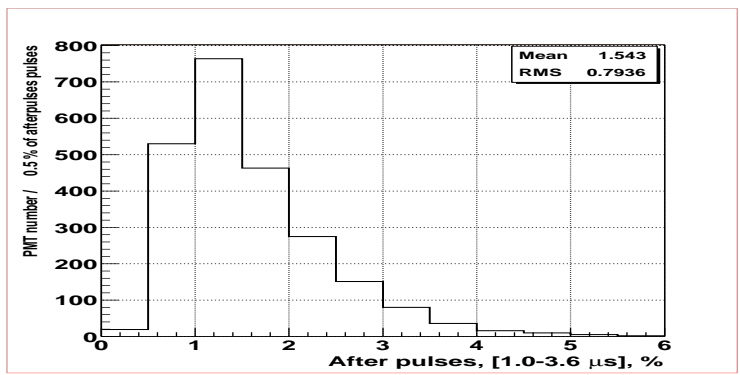}\end{center}

\caption{\label{Fig:ap2}Afterpulses in the region $1.0-3.6$ $\mu$s,}
\end{figure*}

\begin{figure*}
\begin{center}\includegraphics[%
  width=0.60\textwidth,
  height=0.30\textwidth]{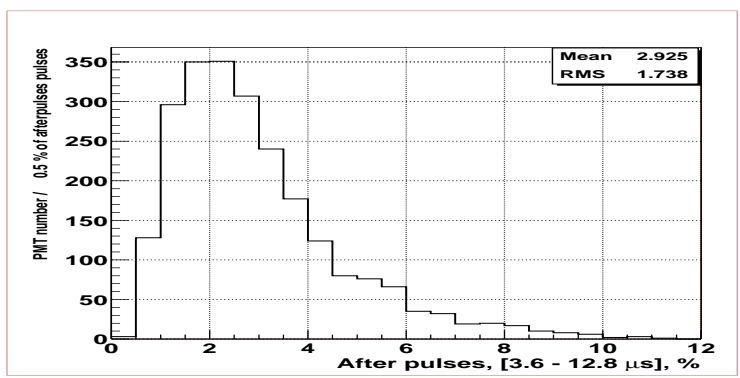}\end{center}

\caption{\label{Fig:ap3}Afterpulses in the region $3.6-12.8$ $\mu$s,}
\end{figure*}

\begin{figure*}
\begin{center}\includegraphics[%
  width=0.60\textwidth,
  height=0.30\textwidth]{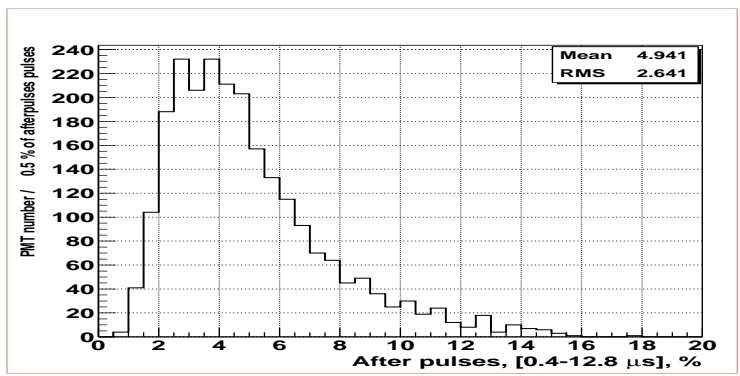}\end{center}

\caption{\label{Fig:apTot}Afterpulses in the region $0.4-12.8$ $\mu$s,}
\end{figure*}

For studying the afterpulses 3 regions of interest were chosen: $0.4-1.0$
$\mu$s, $1.0-3.6$ $\mu$s and $3.6-12.8$ $\mu$s. The amount of
the afterpulses in these regions and the total amount of afterpulses
are presented in Fig.\ref{Fig:ap1}-\ref{Fig:apTot}. The mean (total)
count of afterpulses calculated from the Fig.\ref{Fig:AverageAP}
is $4.9\%$ (see Fig.\ref{Fig:apTot} as well).

The mean excess rate of the afterpulses and the average peak positions
in each of 3 regions is presented in Table \ref{Table:Afterpulses}.
The total afterpulse rate measured in the test facility is somewhat
greater than the one reported by the manufacturer. Indeed, we found
a factor 2 discrepancy. The possible reason could be the different
regime of the PMT operation, in our case the PMT are operating at
a gain of $2\times10^{7}$ in comparison with the gain of $1\times10^{7}$
of the regime recommended by manufacturer. Another possible reason
could be the different discriminator threshold setting. The measurements
with a different threshold showed that the amount of afterpulses decreases
with a higher threshold. In particular, with a threshold set at 0.4
p.e. level one obtains practically the same values as provided by
ETL. 

\begin{table*}
\begin{tabular}{|c|c|c|c|c|}
\hline 
region ($\mu$s)&
$0.4-1.0$&
$1.0-3.6$&
$3.6-12.8$&
$12.8-25$\tabularnewline
\hline 
Total amount, \%&
0.5&
1.5&
2.9&
0.1\tabularnewline
\hline 
average peak position, $\mu$s&
0.88&
1.74&
6.38&
not defined\tabularnewline
\hline 
standard deviation from &
0.09&
0.22&
0.30&
\tabularnewline
the average position, $\mu$s&
&
&
&
\tabularnewline
\hline 
mean excess rate in the peak, Kcps&
21.7&
16.7&
15.4&
3.8\tabularnewline
\hline 
standard deviation from the &
13.6&
7.9&
7.8&
2.7\tabularnewline
mean excess rate, Kcps&
&
&
&
\tabularnewline
\hline
\end{tabular}

\caption{\label{Table:Afterpulses}Main characteristics of the afterpulses.}
\end{table*}

\section{Concluding remarks}

More than 150 PMTs were rejected during the acceptance tests of 2350
PMTs delivered from the manufacturer. The main reasons were the high
dark rate (58 PMTs) and the high afterpulse rate (58 PMTs). 23 PMTs
had a bad single electron charge resolution and 13 PMTs had a bad
time resolution. The data is summed in Table \ref{Table:Rejected}.
The selected 2200 PMTs, that met the requirements of the Borexino
detector, were equipped with light concentrators and $\mu$-metal
screens and installed in the detector. At the moment of the article
submission (June, 2004), all the PMTs are installed in the Borexino
detector, the detector is sealed and ready to be filled with water.

\begin{table*}
\begin{center}\begin{tabular}{|c|c|c|c|c|c|}
\hline 
&
&
\multicolumn{4}{c|}{The reason of rejection}\tabularnewline
\hline 
&
total &
high&
high&
low charge &
low time \tabularnewline
&
rejected&
dark rate&
afterpulse rate&
resolution&
resolution\tabularnewline
\hline
\hline 
Total number&
152&
58&
58&
23&
13\tabularnewline
\hline 
Percentage of the rejected&
100&
38&
38&
15&
9\tabularnewline
\hline 
Percentage of the tested&
6.8&
2.6&
2.6&
1&
0.6\tabularnewline
\hline
\end{tabular}\end{center}

\caption{\label{Table:Rejected}The results of the acceptance test}
\end{table*}

\section{Acknowledgements}

We are deeply grateful to G.Korga and L.Papp who took an active part
in the test. We also thank the LNGS staff for the warm atmosphere
and the good working conditions. The job of one of us (O.S.) was supported
by the INFN sez. di Milano, and he is personally indebted to Prof.
G.Bellini for the possibility to work at the LNGS laboratory. The
authors appreciate the help of M.Laubenstein in preparation of the
manuscript.

Credit is given to the developers of the CERN ROOT program \cite{ROOT},
that was used in the calculations and to create all the figures of
the article.

\end{document}